\documentclass{pasj00}
\draft

\begin{document}

\SetRunningHead{Hillier et al.}{Plasma Blob Ejection from a Quiescent Prominence}

\title{Observations of Plasma Blob Ejection from a Quiescent Prominence by Hinode SOT}

\author{Andrew \textsc{Hillier}\altaffilmark{1}, Hiroaki \textsc{Isobe}\altaffilmark{1,2} and Hiroko \textsc{Watanabe}\altaffilmark{1}}
\altaffiltext{1}{Kwasan and Hida Observatories, Kyoto University, Yamashina-ku, Kyoto, 607-8417, Japan}
\altaffiltext{2}{Unit of Synergetic Studies for Space, Kyoto University, Sakyo-ku, Kyoto 606-8502, Japan}
\email{andrew@kwasan.kyoto-u.ac.jp}

%

\KeyWords{Sun: prominences Sun: magnetic fields} 

\maketitle

\begin{abstract}
We report findings from 0.2'' resolution observations of the 2007 October 03 quiescent prominence observed with the Solar Optical Telescope on the Hinode satellite.
The observations show clear ejections from the top of the quiescent prominence of plasma blobs.
The ejections, originating from the top of rising prominence threads, are impulsively accelerated to Alfv\'enic velocities and then undergo ballistic motion.
The ejections have a characteristic size between $ \sim$ 1000\,--\,2000\,km.
These characteristics are similar to downwardly propagating knots (typical size $ \sim 700$ km) that have been observed in prominence threads, we suggest that the plasma blob ejections could be the upward moving counterpart to the downwardly propagating knots.
We discuss the tearing instability as a possible mechanism to explain the ejections.
\end{abstract}

\section{Introduction}

Quiescent prominences are large structures composed of relatively cool plasma, that exist in quiet-sun regions of the corona. 
Globally, quiescent prominences are incredibly stable, often existing in the corona for weeks.
It is known that the temperature of quiescent prominences is about 10000 K (\cite{TH1995}) and density is $\sim 10^{-11}$ g cm$^{-3}$ (\cite{HIR1986}) which is a decrease and increase of two orders of magnitude respectively from the surrounding corona. 
Using this value for the temperature, it can be calculated that the pressure scale height ($\Lambda $) of a quiescent prominence is $\Lambda \approx 300km$, which is more than 1 order of magnitude less than the characteristic height of a quiescent prominence ($\approx 25$Mm)
Using a characteristic gas pressure of $0.6~dyn~cm^{-2}$ (\cite{HIR1986}) and magnetic field of $3 - 30$ G (\cite{LER1989}), gives a plasma $\beta \sim 0.01$\,-\,$0.1$. 

In contrast to this global stability, it has long been known that locally quiescent prominences are highly dynamic phenomena.
There are reports of downflows (\cite{Eng1981}),vorticies of approximately $10^5$ km $\times 10^5$ km in size (\cite{LZ1984}) and a bubble of size $2800$ km forming a keyhole shape with a bright centre (\cite{DT2008}) in quiescent prominences.
The motion of these phenomena was found to be 10-30\,km s$^{-1}$.

Observations of quiescent prominences by the Solar Optical Telescope (SOT) on Hinode (\cite{TSU2007}, \cite{KOS2007}), with 0.2 {\arcsec} spatial resolution in seeing free conditions, have greatly extended our knowledge of prominence dynamics through the discovery of new dynamic phenomena.
\citet{Ber2010} reported turbulent upflows that propagated from cavities that formed at the bottom of prominences through a height of about 10Mm.
Bright descending knots that form inside prominence threads and travel at an average speed of $16$ km s$^{-1}$ were observed by \citet{Chae2010}.
The observed knots were impulsively accelerated throughout their lifetimes.
\citet{ZAP2007} described the 3D velocity of ``blobs'' ejected from active region prominences and found that there was a significant line of sight velocity component associated with the ejections.

In this paper, we report the observation performed on 2007 October 3.
A quiescent prominence was seen on the NW solar limb, and many downward propagating knots are observed.
During the course of these observations we found upwardly ejected plasma blobs observed with Hinode SOT.
The high spatial and temporal resolution of the images allows the motion of the plasma blobs to be accurately determined.
This paper is structured as follows, in section 2 the observational data will be presented, section 3 contains the analysis of the data and a discussion of the impacts of these results is presented in section 4.


\section{Observation}

\begin{figure*}[ht]
  \begin{center}
    \FigureFile(120mm,120mm){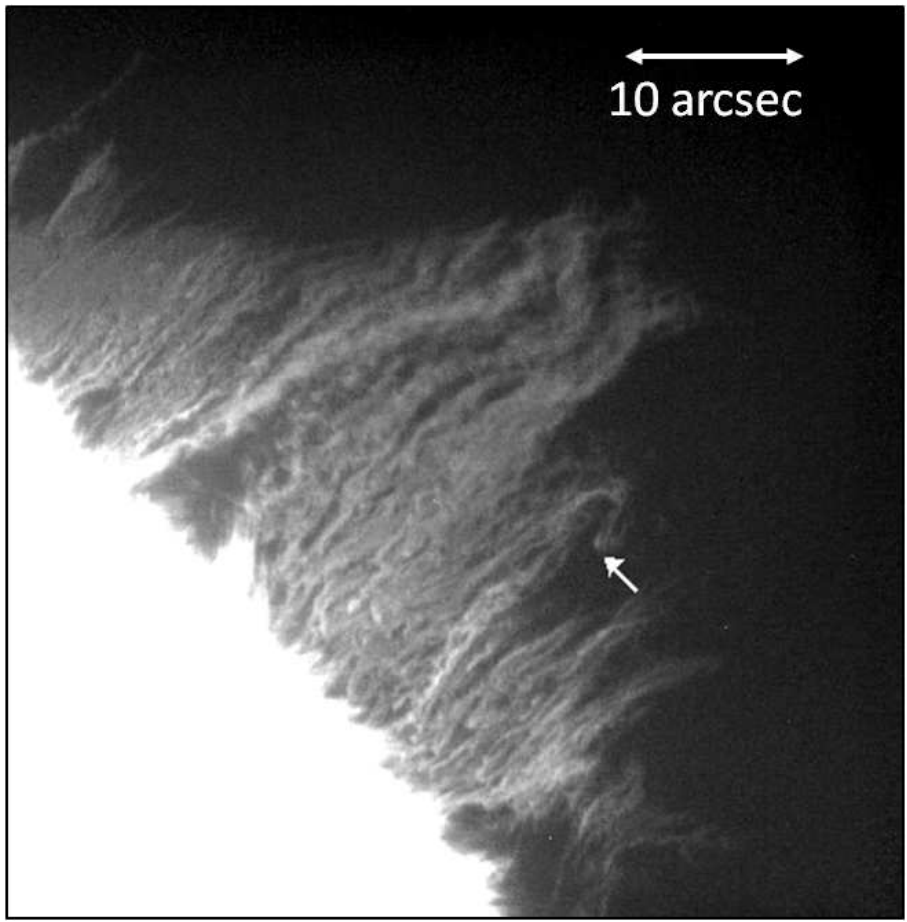}
  \end{center}
  \caption{Quiescent Prominence observed in the Ca $_{II} $ H-line 396.8 nm spectral line on 2007 October 03 04:52UT. Disk position 41 N 84 W. The pixel size is 0.108 arcsec pixel$^{-1}$. The arrow denotes the position of ejection. The intensity is saturated to show the off limb prominence.}\label{fig1}
\end{figure*}

Figure \ref{fig1}, shows a quiescent prominence seen on the NW solar limb (41$^{\circ}$N 84$^{\circ}$W) on 2007 October 03 observed by the SOT with the Ca {\footnotesize II} H filter at a cadence of 30 s.
The time series of this observation was between 01:16UT and 04:59UT.
This prominence presents many interesting dynamic features, for example the start of this observation (01:16UT) a clear cavity has formed inside the prominence similar to those described in \citet{Ber2010}.
There are also a number of bright threads and downwardly propagating knots that occur during the duration of the observations.

\section{Analysis of Data}

This paper focuses on the occurrence of plasma blob ejections, that are ejected against gravity, from the top of the quiescent prominence.
One such example, denoted P3, is highlighted by the arrow in Figure \ref{fig1}.
During the 3.75h observation, four such plasma blobs were observed.
Eye detection was used to determine the occurrence of plasma blob ejection, therefore the number of plasma blobs found should be viewed as the minimum number of plasma blobs that were ejected.
To determine the physical parameters of the plasma blobs, the position is first determined by eye detection, this position is then refined by employing a brightest pixel detection routine in a $21 \times 21$ pixel ($1470$ km$\times 1470$ km) square.
This point is considered to be the center of the plasma blob.
The newly determined central point of the plasma blob is used to perform a Gaussian fit of the intensity profile along the X-direction through the center of the plasma blob.
The full width half maximum of the Gaussian curve is considered to be the size of the plasma blob.

\begin{figure*}[ht]
  \begin{center}
    \FigureFile(160mm,140mm){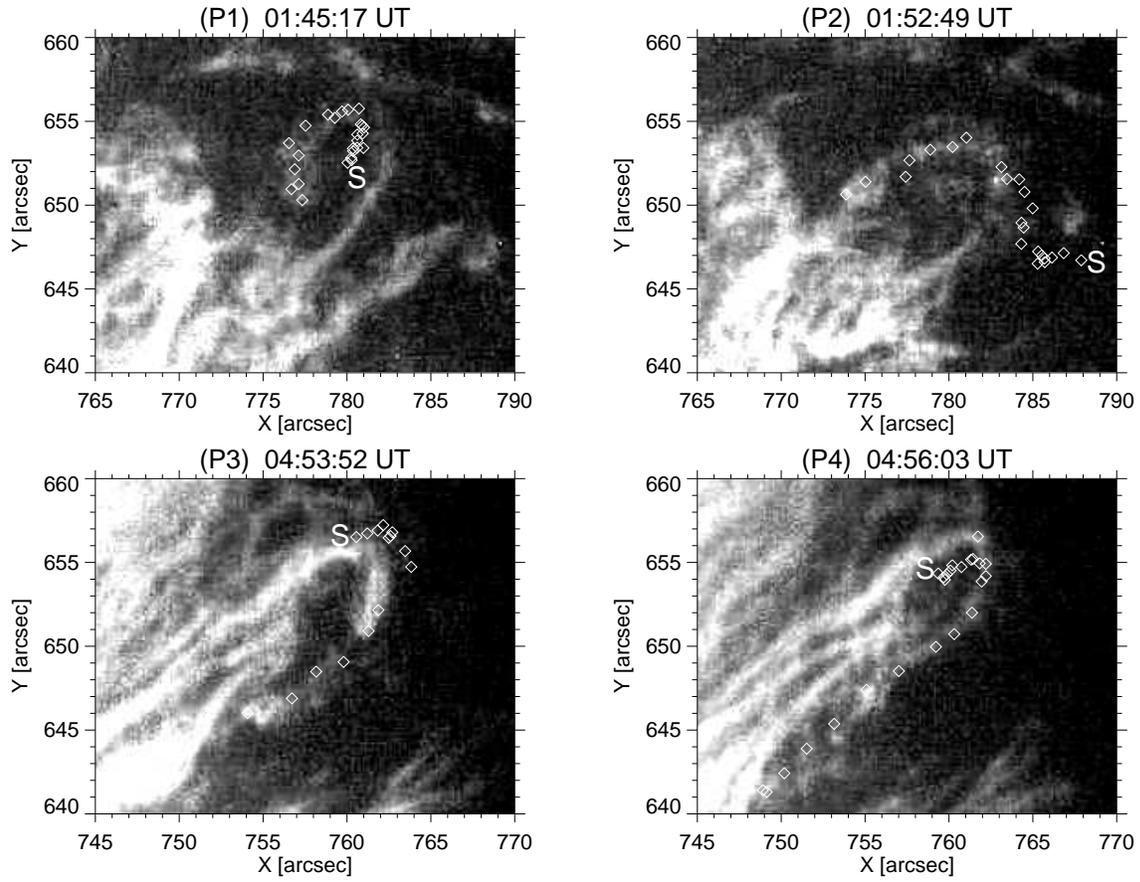}
  \end{center}
  \caption{Evolution of plasma blob ejection for P1, P2, P3 and P4. Diamonds show temporal evolution of plasma blob, where the start position is mark by `s'. The background image shows the final image of the plasma blob.}\label{fig2}
\end{figure*}

The ejected plasma blobs (denoted P1, P2, P3 and P4) have characteristic size $1000 - 2200$ km with maximum horizontal velocity (velocity, in the plane of sky, perpendicular to gravity) between $27 $\,-\,$ 42$ km s$^{-1}$ and maximum vertical velocity (velocity against gravity) between $6$\,-\,$30$ km s$^{-1}$. 
The parameters for all plasma blobs are displayed in Table \ref{table1}.
The error in these velocity measurements is calculated from the error in calculation the position, roughly estimated from Figure \ref{fig2} to be up to 5\,px (=350\,km), corresponding to an apparent velocity error up to 11\,km\,s$^{-1}$. 

Figure \ref{fig2} shows a snapshot of each plasma blob, where the diamonds show the movement of the plasma blob with the initial position marked by ``S''.
The characteristic evolution can be described as follows.
First a prominence thread slowly rises up with a velocity of a few km s$^{-1}$.
The plasma blob is then ejected, reaching Alfv\'enic velocity often with a very strong horizontal velocity component.
The ejected plasma blob then undergoes ballistic motion.
P1 and P3 show a change in the direction of horizontal motion.

\begin{figure*}[ht]
  \begin{center}
   \FigureFile(120mm,120mm){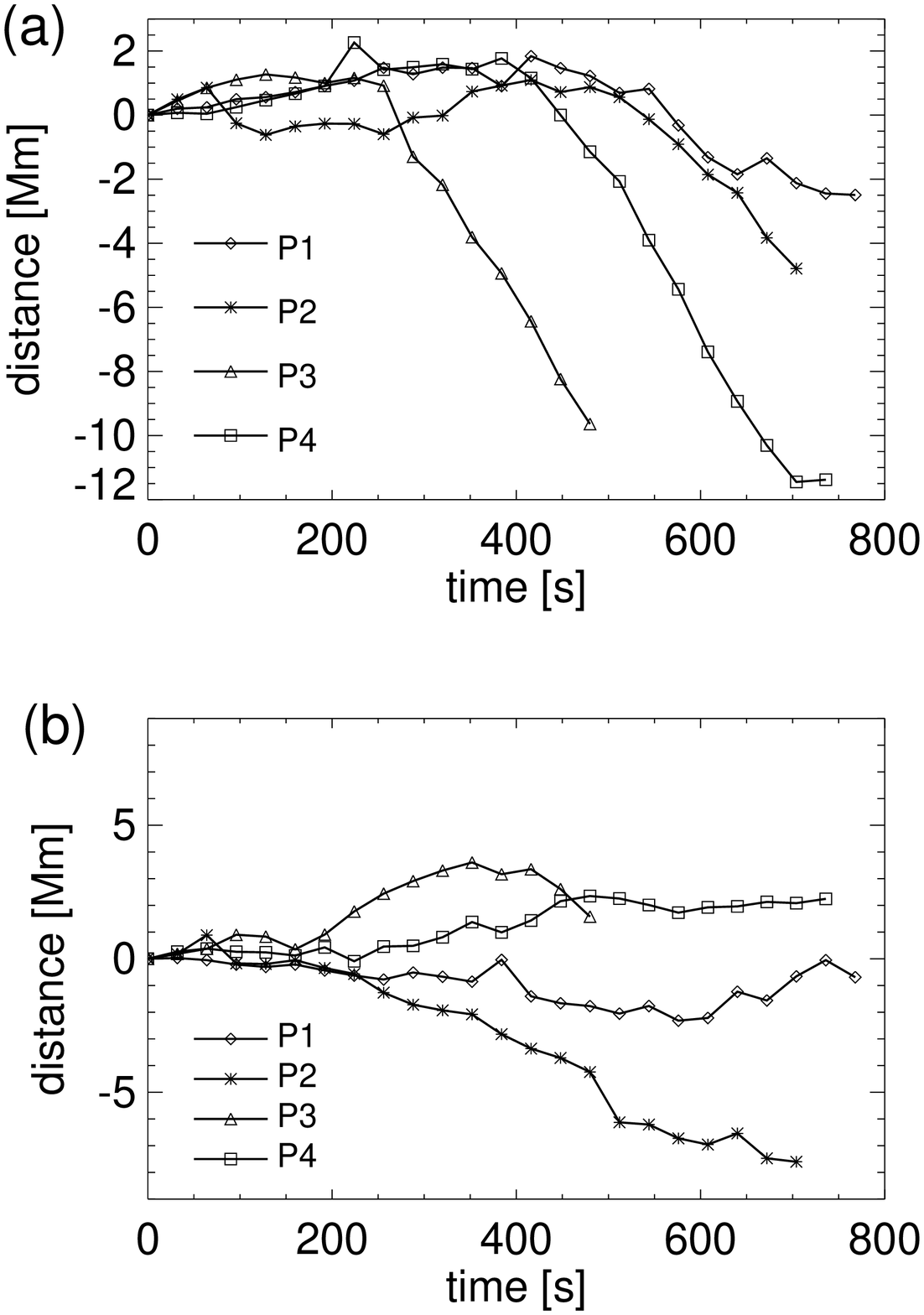}
  \end{center}
  \caption{Evolution of (a) vertical and (b) horizontal position of plasma blob}\label{fig3}
\end{figure*}

\begin{figure*}[ht]
 \begin{center}
    \FigureFile(150mm,150mm){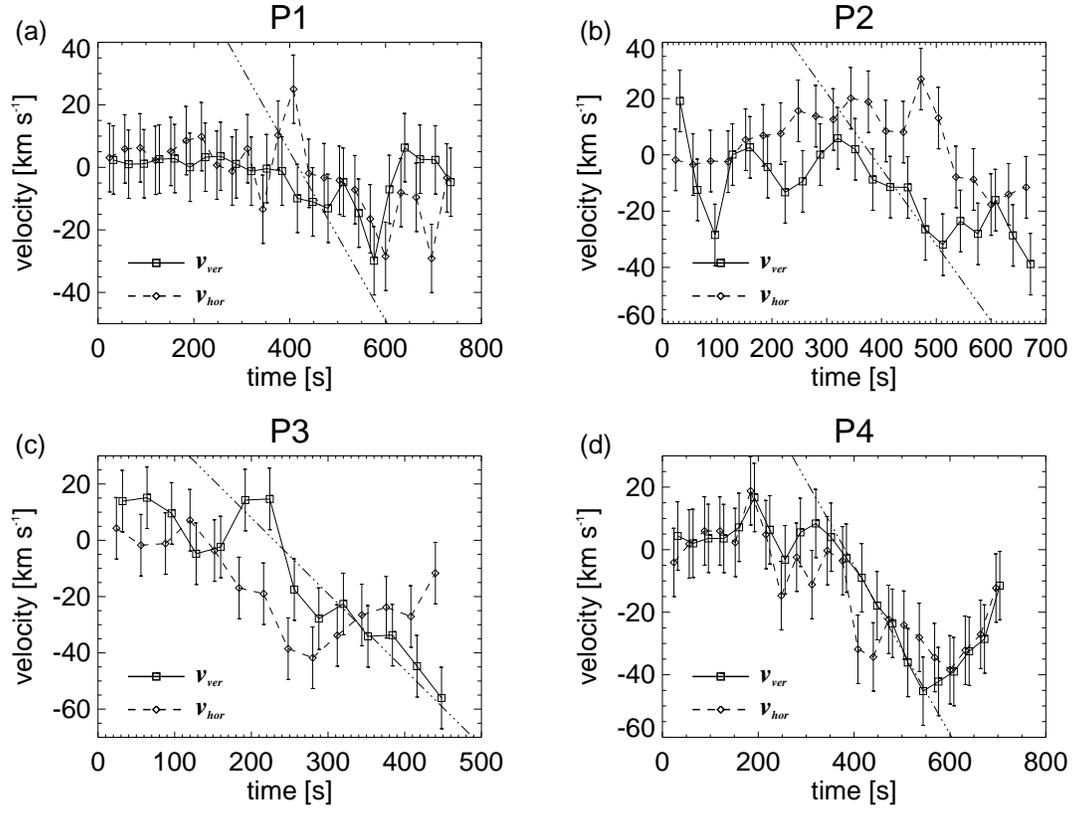}
  \end{center}
  \caption{Evolution of vertical and horizontal velocity of plasma blob (a) P1, (b) P2, (c) P3 and (d) P4. The dash and three dots line shows the slope given by acceleration due to gravity. The error bars are calculated from the position detection error of up to 5\,px (=350km), roughly estimated from Figure \ref{fig2} corresponding to an apparent velocity error up to 11\,km\,s$^{-1}$.}\label{fig4}
\end{figure*}

Figure \ref{fig3} shows the change in position (both (a) vertical and (b) horizontal) for each plasma blob.
The position $(0,0)$ is the position of the plasma blob at the start time given in table \ref{table1}.
The evolution shows the three phases phases described above: slow rise, impulsive acceleration and free fall under constant gravity.
This can also be seen in Figure \ref{fig4} where the vertical and horizontal component of the velocity, with respect to gravity, is shown for each plasma blob.
The velocity at time $t_n$ is calculated using the slope given by a linear fitting of the position at time $t_{n-1}$, $t_n$ and $t_{n+1}$. 
The slope of the dash and three dot line shows free fall under gravitational acceleration at the solar surface of $2.7 \times 10^2$ m s$^{-2}$.

The three stages of the evolution will now be described individually using the evolution of P2 as an example.
Initially, the position of the plasma blob (thread top) remains relatively motionless.
Around $t=352$s, the plasma blob develops a strong horizontal velocity ($\approx 20$ km s$^{-1}$).
At this time the vertical velocity is approximately $\approx 2$ km s$^{-1}$.
After this time the plasma blob appears to undergo ballistic motion for approximately $200$ s.
At approximately 550\,s the horizontal motion is quenched.

In the evolution of P1, after about $600$ s the vertical position becomes almost constant with a velocity that is only a few km s$^{-1}$.
In place of this, the plasma blob changes horizontal direction developing a strong horizontal velocity of approximately $20$ km s $^{-1}$.
This velocity is approximately the value of the velocity associated with the vertical decent of the plasma blob before $600$ s.
The evolution of P3 also shows a change in the horizontal direction the plasma blob is traveling in.


\section{Discussion}

The observations show clear ejections of prominence material into the corona.
The properties of these ejections can be summarized as:
\begin{enumerate}
\item Plasma blobs are ejected, with a component of the velocity that is against gravity, from the top of the quiescent prominence at a height of approximately 25 Mm.

\item The plasma blob motion has three components: slow rise phase,  impulsive acceleration to Alfv\'enic velocities and ballistic motion.  

\item The plasma blobs have characteristic size of $\sim 1000 - 2000$ km.

\item Plasma blob ejection activity seemed to be  concentrated spatially, with all plasma blobs being ejected in 15 Mm of each other, P1 and P2 from within 5 Mm and P3 and P4 from the same thread. This may be a result of the selection method.

\end{enumerate}

The impulsive acceleration to Alfv\'enic velocities implies that the plasma blob ejections are most likely to have been caused by magnetic reconnection.
Once the plasma blobs are ejected they then undergo ballistic motion until they fall to a point where they become no longer visible against the background prominence material.
The final velocity of P1 is almost purely horizontal, this implies that the plasma blob is following the direction of the magnetic field.

The tearing instability (a resistive instability where reconnection creates magnetic islands) provides a possible explanation for these observations of ejections of plasma blobs.
Current sheets formed inside the prominence, either the current sheet that a prominence thread uses to support the prominence material or a current sheet that forms between the magnetic field of two different threads, are potential sites for the tearing instability to occur.
Once the tearing instability has been excited, plasmoids are formed and then ejected from the current sheet after going through coalescence.
The tearing instability would create plasmoids that are ejected both with and against gravity.
The plasmoid would then undergo ballistic motion as the main force acting on the plasmoid would be gravity.
The knots described in \citet{Chae2010}, of similar size and velocity, could be the downwardly propagating counterpart of this process. 
\citet{HILL2010} described how the prominence geometry could be affected by the partially ionized prominence material, and in turn how this could increase the growth rate of the tearing instability.

The proposed method for knot formation in \citet{Chae2010} would be able to explain the creation of upward ejections through the release of magnetic tension.
However, the ejections described in this paper are initiated with a strong horizontal component of the velocity, then follow ballistic motion and appear to travel along the direction of the magnetic field.
All the ejections end their observable motion at a position closer to the solar surface than the point they are ejected.
Therefore, assuming that on dynamic timescales the plasma behaves in an ideal MHD fashion, this model would imply there is ultimately an increase in magnetic tension instead of a release.
There also does not seem to be a way of accelerating the plasma blob in the direction of the magnetic field.

However these plasma blobs are formed, they provide another example of flows that are created inside prominences which move against gravity.
Using the fact that plasma $\beta <1$ and the the velocities are Alfv\'enic, then it can be assumed that the plasma blobs ejections presented in this paper are of a magnetic origin.
For the magnetic field to drive such ejections it is necessary that locally a significant change in the structure of the magnetic field occurs.
How this local change in structure can be match with the global stability is still not understood.

Further work would be necessary to give conclusive answers as to if the tearing instability is the root of these observations, and to explain why the downwardly propagating knots appear with greater frequency.
If the tearing instability is proved to be an effective process in quiescent prominences, this could lead to the creation of a tangled magnetic field as proposed by \citet{VB2010} and may be able to explain the complex structure of prominences.


\begin{table*}
  \caption{Parameters of plasma blob ejections. Av. I is the average intensity of the plasma blob, $v_{ver}$ is the velocity against gravity and $v_{hor}$ is the velocity perpendicular to gravity}\label{table1}
  \begin{center}
    \begin{tabular}{llllllll}
      \hline
      No. & Av. I & size (km) & time (UT) & Max $v_{ver}$  (km s$^{-1}$) & Min $v_{ver}$ (km s$^{-1}$) & Max $|v_{hor}|$  (km s$^{-1}$)\\
      \hline
      P1 & 38 & 1000 & 01:35 & 6 & -30 &  29 \\
      P2 & 36 & 2100 & 01:50 & 15 & -39 &  27 \\
      P3 & 42 & 1500 & 04:50 & 17 & -55 &  42 \\
      P4 & 44 & 2200 & 04:50 & 30 & -45 &  38 \\
      \hline
    \end{tabular}
  \end{center}
\end{table*}

\bigskip

Hinode is a Japanese mission developed and launched by ISAS/JAXA, with NAOJ as domestic partner and NASA and STFC (UK) as international partners. It is operated by these agencies in co-operation with ESA and NSC (Norway).
The authors would like to thank the staff and students of Kwasan and Hida observatories for their support and comments, special mention goes to K. Shibata for the useful insights provided.
This work was supported in part by the Grant-in-Aid for the Global COE program ``The Next Generation of Physics, Spun from Universality and Emergence'' from the Ministry of Education, Culture, Sports, Science and Technology (MEXT) of Japan.
AH is supported in part by the Japanese government (Monbukagakusho) scholarship from the Ministry of Education, Culture, Sports, Science and Technology (MEXT) of Japan. 
HI is supported by the Grant-in-Aid for Young Scientists (B, 22740121).


\end{document}